\documentclass{llncs}
\usepackage{url}

\begin{document}
\mainmatter

\title{Optimizing Affine Maximizer Auctions via Linear Programming: an Application
to Revenue Maximizing Mechanism Design for Zero-Day Exploits Markets}

\author{Mingyu Guo\inst{1} \and Hideaki Hata\inst{2} \and Ali Babar\inst{1}}
\institute{
    School of Computer Science\\University of Adelaide, Australia\\
    \{mingyu.guo, ali.babar\}@adelaide.edu.au
    \and
    Graduate School of Information Science\\Nara Institute of Science
    Technology, Japan\\
    hata@is.naist.jp}

\maketitle

\begin{abstract} Optimizing within the affine maximizer auctions (AMA) is an
    effective approach for revenue maximizing mechanism design. The AMA
    mechanisms are strategy-proof and individually rational (if the agents'
    valuations for the outcomes are nonnegative). Every AMA mechanism is
    characterized by a list of parameters. By focusing on the AMA mechanisms,
    we turn mechanism design into a value optimization problem, where we only
    need to adjust the parameters. We propose a linear programming based
    heuristic for optimizing within the AMA family. We apply our technique to
    revenue maximizing mechanism design for zero-day exploit markets. We show
    that due to the nature of the zero-day exploit markets, if there are only
    two agents (one offender and one defender), then our technique generally
    produces a near optimal mechanism: the mechanism's expected revenue is
    close to the optimal revenue achieved by the optimal strategy-proof and
    individually rational mechanism (not necessarily an AMA mechanism).
\end{abstract}


\section{Introduction}

Revenue maximizing mechanism design is a fundamental topic in algorithmic game
theory. Myerson~\cite{Myerson81:Optimal} solved for the revenue maximizing
mechanism for selling a single item, subject to a technical condition called
the {\em
monotone hazard rate} condition. Myerson's optimal auction is
surprisingly elegant.
For example, if every agent's type is drawn from an identical and
independent distribution, then the optimal mechanism is simply the Vickrey
auction~\cite{Vickrey61} with a reserve price. Unfortunately, Myerson's
technique does not generalize to more complex settings. For example, when it
comes to combinatorial auctions (auctions where multiple items are for sale,
and the agents bid on bundles of items), revenue maximizing mechanism design
remains an open problem. Another notable application domain of revenue
maximizing mechanism design is the sponsored search
auctions~\cite{Lahaie07:Sponsored}, where the search engines sell advertisement
slots to advertisers, aiming to maximize revenue besides other objectives.
Even though no optimal mechanisms have been derived for general combinatorial
auctions or sponsored search auctions, for restricted domains, well performing
mechanisms have been obtained based on a variety of revenue-boosting
techniques~\cite{Emek12:Signaling,Guo13:Revenue,Guo14:Increasing,Likhodedov04:Boosting,Likhodedov05:Approximating}.

There are several general revenue-boosting techniques. For example, we may
artificially increase the winning chance of lower bidders, in order to drive up
the competition faced by the higher bidders. As another example, we may
artificially discourage or outright ban certain outcomes, in order to prevent
low-revenue outcomes or force the agents to pay more to achieve the discouraged
outcomes. The above techniques form the basis of a family of mechanisms called
the {\em affine maximizer auctions (AMA)}. Lavi {\em et
al.}~\cite{Lavi03:Towards} conjectured that a combinatorial auction is truthful
if and only if it is an AMA mechanism, subject to technical conditions.
Likhodedov and Sandholm~\cite{Likhodedov04:Boosting,Likhodedov05:Approximating}
studied revenue maximizing combinatorial auction design by optimizing within
the family of AMA mechanisms. The idea of optimizing within the AMA family is a
general approach that can be applied to many different mechanism design
settings, because generally the AMA mechanisms are well defined and the family
contains a large number of mechanisms. By optimizing within the AMA family,
there is a good chance of reaching a well-performing mechanism in terms of
revenue. However, the issue with optimizing within the AMA family is that every
AMA mechanism is characterized by $|O|+n$ parameters, where $|O|$ is the size
of the outcome space $O$, and $n$ is the number of agents. For combinatorial
auctions, $|O|$ is exponential in the number of items, which makes it
computationally impractical to optimize within the AMA family for this setting.
Due to this, Likhodedov and Sandholm only studied the AMA family for the case
of selling two items. When there are only two items, the number of parameters
is small enough for the authors to conduct optimization via grid-based gradient
descent.\footnote{The authors also proposed a restricted version of AMA called
    the VVCA mechanisms. A VVCA mechanism is only characterized by $2n$
    parameters, which makes it much easier to optimize over. On the other hand,
due to the fact that the VVCA family is only a tiny subset of the whole AMA
family, we lose revenue by focusing only on it.}

In this paper, we propose a linear programming based technique for optimizing
within the AMA family. Every outcome corresponds to one variable in our LP
model. As a result, our technique can handle reasonably large number of
outcomes. For example, let us consider the case where $|O|$ is a few hundred.
Running a LP with a few hundred variables is computationally tractable. On the other hand, methods
such as grid-based gradient descent are impractical.

We apply our new technique to a specific mechanism design problem. Our paper
focuses on revenue maximizing mechanism design for zero-day exploit
markets~\cite{Guo16:Revenue}. Zero-day exploits refer to software
vulnerabilities that are not known to the software vendor. Trading zero-day
exploits as legitimate business is a recent trend in the security
industry~\cite{Egelman13:Markets}. According to a price list collected by
Greenberg~\cite{greenberg2012forbes}, the price of a zero-day exploit is
between \$5000 to \$250,000. There are venture capital backed security
consulting companies whose business model is selling zero-day
exploits~\cite{fisher2015threatpost}. One of the companies mentioned in
\cite{fisher2015threatpost} even offered one point five million US dollars for one new iOS exploit. The
reason an exploit can be priced so high is that generally it can stay
alive for a long period of time~\cite{Bilge12:Before}. Unless the software
vendor is informed about an exploit, there is very low chance for an exploit
to be discovered independently. To remedy this, software vendors often run bug
bounty programs, which are markets where the software vendors buy exploits from
security researchers~\cite{algarni2014software,guideline2015chrome}.

Guo {\em et al.}~\cite{Guo16:Revenue} proposed a formal mechanism design model
for zero-day exploit markets. In the authors' model, one exploit is being sold
to multiple buyers over a period of time $[0,1]$. The model is different from the classic single-item auction
for the following reasons:

\begin{itemize}
    \item There are two categories of buyers. The {\em defenders} buy exploits
        to fix them. Typically there is only one defender, which is the
        software vendor. The {\em offenders} buy exploits to utilize them.
        National security agencies and police are example offenders. For
        example, Zerodium~\cite{fisher2015threatpost} is a consulting company
        that buys zero-day exploits and resells them to mostly government
        agencies. The offenders wish to utilize an exploit for as long as
        possible. Once an exploit is obtained by a defender, the exploit
        becomes worthless. Using mechanism design terminologies, the buyers have
        {\em externalities}.\\

    \item The item being sold is an informational item, which means that we can
        sell the same item multiple times ({\em e.g.}, to multiple offenders).
        Of course, once an exploit is sold
        to a defender, we cannot sell it to any offenders {\em afterwards},
        because it has become worthless.
        \\

    \item Because the item being sold is a piece of information, we cannot
        simply describe it in full details to the buyers without some kind of payment enforcing
        mechanism, because otherwise the buyers can walk away with the exploit for
        free. Furthermore, we cannot ask the buyers to bid on an exploit that
        carries no description, because the buyers cannot come up with their
        private valuations if no description is given.
\end{itemize}

Guo {\em et al.}~\cite{Guo16:Revenue} proposed a mechanism property called
{\em straight-forwardness}: a mechanism is straight-forward if it describes the
exploit in full details to the offenders, before they submit their bids. This
is required because typically offenders already have many exploits in their
arsenals. With full description, they can evaluate whether the exploit being
sold is original, and to what extent the exploit helps them.
Straight-forwardness does not require the mechanism to describe the exploit to
the defenders before they bid (otherwise, the exploit gets fixed).
Straight-forwardness only describes to the defenders how severe the exploit is:
{\em e.g.}, this exploit allows anyone to remotely control an iOS device. From
the perspective of a defender, every exploit is new (otherwise, it wouldn't be
an exploit). We ask the defenders to come up with their private valuations
based on the exploit's severity, which is exactly how bug bounty markets operate (in
bug bounty markets, bugs are priced according to their severity
levels~\cite{guideline2015chrome}).

Guo {\em et al.}~\cite{Guo16:Revenue} showed that if straightforwardness is
required together with strategy-proofness and individual rationality, then one
revenue-maximizing mechanism must work as follows: we describe the exploit in
full details to all offenders at time $0$. We also describe the exploit's
severity level to the defenders at time $0$. The offenders and defenders submit
their bids. The offenders bid to keep the exploit alive for as long as
possible. The defenders bid to kill off the exploit as early as possible. In
some sense, the model is similar to the {\em cake-cutting}
problem~\cite{Brams07:Better,Chen10:Truth} and the {\em single facility
location}
problem~\cite{Goemans04:Cooperative,Procaccia09:Approximate}.\footnote{In our
    model, we allow payments. After all, the objective is to maximize revenue.}
For this model, the authors proposed one heuristic-based randomized AMA
mechanism.

In this paper, for the above model, we use our new technique to optimize within
the AMA family. We also show that if there are only two agents (one offender
and one defender), and if the defender's valuation is much lower than the
offender's (typically true for zero-day exploit markets), the optimal AMA
mechanism's revenue is close to the optimal revenue. For demonstrating this result,
we propose and study a family of mechanisms called the {\em posted-price}
mechanisms for our model.

\section{Model Description}

We use $O$ to denote the outcome space. We use $\Theta_i$ to denote agent $i$'s
type space. We use $v_i(\theta_i,o)$ to denote agent $i$'s valuation for
outcome $o\in O$ when her type is $\theta_i\in \Theta_i$.

For the zero-day exploit mechanism design model proposed in Guo {\em et
al.}~\cite{Guo16:Revenue}, the outcome space is $[0,1]$. An outcome $o \in
[0,1]$ represents when the exploit is killed off (revealed to the
defenders).\footnote{If we allow randomized mechanisms, then an outcome
    is a nonincreasing function $o(t)$, with $o(0)=1$ and $o(1)=0$. $o(t)$
represents the probability for the exploit to be alive at time $t$.}
In order to run the technique proposed in this paper, we require the outcome
space to be finite. So for this technical reason, we set the outcome space to
be $\{0,\frac{1}{k},\frac{2}{k},\ldots,1\}$. That is, we will only reveal the
exploit at these discrete moments. The size of the outcome space $|O|=k+1$.

The family of AMA mechanisms is defined as follows:

\begin{itemize}
    \item Given a type profile $\theta$, the outcome picked is the following:
        \[o^*=\arg\max_{o\in O}\left(\sum_{i=1}^nu_iv_i(\theta_i,o)+a_o\right)\]

    \item Agent $i$'s payment equals:
        \[\frac{\max_{o\in O}\left(\sum_{j\neq i}u_jv_j(\theta_j,o)+a_o\right)
        - \sum_{j\neq i}u_jv_j(\theta_j,o^*)-a_{o^*}}{u_i}\]
    \end{itemize}

In the above description, the $u_i$ and the $a_o$ are constant parameters.
$u_i\ge 1$ for all $i$. The $a_o$ are unrestricted. In total, there are $n+|O|$
parameters. Every AMA mechanism is characterized by these many parameters. For
any assignments of the parameters, the
corresponding AMA mechanism is strategy-proof. However, not every AMA mechanism
is individually rational. If we further assume that $\forall i,\theta_i,o$,
$v_i(\theta_i,o)\ge 0$, then every AMA mechanism is individually rational. To
show this, we only need to show that an agent's valuation is always at least
her payment. That is, \[v_i(\theta_i,o^*)\ge \frac{\max_{o\in
O}\left(\sum_{j\neq i}u_jv_j(\theta_j,o)+a_o\right) - \sum_{j\neq
i}u_jv_j(\theta_j,o^*)-a_{o^*}}{u_i}\] \[\iff
\sum_ju_jv_j(\theta_j,o^*)+a_{o^*}\ge \max_{o\in O}\left(\sum_{j\neq
i}u_jv_j(\theta_j,o)+a_o\right)\] The right-hand side is less than or equal to
the left-hand side if every agent's valuation for every outcome is nonnegative.

In our model, an outcome represents when the exploit is killed off. For
presentation purpose, we sometimes use $t$ to refer to an outcome.

An offender's valuation is defined as:
\[v_i(\theta_i,t) = \int_0^tf_{\theta_i}(x)dx\]

A defender's valuation is defined as:
\[v_i(\theta_i,t) = \int_t^1f_{\theta_i}(x)dx\]

An offender ``enjoys'' the exploit from time $0$ to $t$, and a defender
values the safe period from time $t$ to $1$.
$f_{\theta_i}(x)$ represents agent $i$'s instantaneous value (nonnegative) at time $x$, when
her type is $\theta_i$. Based on the above definitions of the agents' valuations,
we have that every AMA mechanism is individually rational for our model.

\section{Optimizing Affine Maximizer Auctions}

We recall that an AMA mechanism is characterized by $n+|O|$ parameters ($u_i$
for every agent $i$, and $a_o$ for every outcome $o$). For presentation
purpose, we define $Z=n+|O|$ and use $p_1,p_2,\ldots,p_Z$ to refer to the
parameters. Let $M(p_1,p_2,\ldots,p_Z)$ be the AMA mechanism characterized by
$p_1$ to $p_Z$. The task of optimizing within the AMA family is simply to
optimize over the parameters:

\[\max_{p_1,p_2,\ldots,p_Z}ER(M(p_1,p_2,\ldots,p_Z))\]

Here, $ER(M)$ represents mechanism $M$'s expected revenue. We have analytical
characterization of the AMA payments, so the revenue of $M$ given a specific
type profile can be calculated accordingly. Unfortunately, there is no known
short-cut for calculating the expected revenue. Given a prior distribution of
the $\theta_i$, we need to draw large amount of sample profiles to calculate
the expected revenue. For example, if for every agent $i$, we draw $100$
samples for $\theta_i$, then altogether the number of type profiles is $100^n$.
For this reason, in this paper, we focus on cases where $n$ is small.\footnote{We have to emphasize that this is not an uncommon constraint when it comes to
using numerical methods for maximizing mechanism revenue.}

Likhodedov and Sandholm~\cite{Likhodedov04:Boosting,Likhodedov05:Approximating}
used a grid-based gradient descent approach for optimizing the parameters.
Under this approach, suppose we start from a grid point $(p_1,p_2,\ldots,p_Z)$,
we have to examine all neighbouring points $(p_1+\delta_1h, p_2+\delta_2h, \ldots,
p_Z+\delta_Zh)$, where $\delta_i\in \{-1,0,1\}$ and $h$ is the grid size. We
need to examine $3^Z$ points. So this approach requires that both $n$ and $|O|$
be tiny. For example, if $Z=100$, then the approach is impractical. For our
technique, $Z$ is allowed to be large: our technique involves a LP model with $Z$
variables, which takes polynomial time in $Z$.

A high-level description of our optimizing technique is as follows:

\begin{itemize}

    \item We initialize the algorithm with an AMA mechanism: {\em e.g.}, one
        based on random parameters, or the VCG mechanism.

    \item Given $M_0$ characterized by $p_1^0,p_2^0,\ldots,p_Z^0$,
        we use a heuristic to approximate the optimal AMA mechanism near this starting point, using
        a linear program.
        A mechanism $M$ (characterized by $p_1,p_2,\ldots,p_Z$) is near $M_0$
        if $\max_i|p_i-p_i^0|\le \epsilon$ for a threshold $\epsilon$.
        We repeat this step using the new mechanism as the starting point.

\end{itemize}

The above algorithm may end with a locally optimal mechanism, which means
that we may need to repeat the algorithm using different initial points.

Now we present the details of the linear program.
We index the outcomes using $0,1,\ldots,k$.
We denote the initial mechanism as
$M(u_1^0,u_2^0,\ldots,u_n^0,a_0^0,a_1^0,\ldots,a_k^0)$.

The following optimization model solves for the optimal AMA mechanism near this
starting point:

\begin{center}
    \framebox{\parbox{3.5in}{
            {\bf Model 1}\\
            {\bf Variables:} $u_1,u_2,\ldots,u_n,a_0,a_1,\ldots,a_k$\\
            {\bf Maximize:} $ER(M(u_1,u_2,\ldots,u_n,a_0,a_1,\ldots,a_k))$\\
            {\bf Subject to:}\\
            For all $i$, $u_i\ge 1$ and $u_i^0-\epsilon \le u_i\le
            u_i^0+\epsilon$\\
            For all $t$, $a_i^0-\epsilon \le a_i\le a_i^0+\epsilon$
    }}
\end{center}

Of course, the above model is not a linear program, as $ER(M)$ is not a linear
combination of the variables. {\em We will approximate $ER(M)$ using a linear combination
of the variables.}

Let $S$ be a large set of type profiles, we will
approximate $ER(M)$ as follows:

\[ER(M)\approx\sum_{\theta\in S}P(\theta)\sum_iC_i(M,\theta)\]

Here, $C_i(M,\theta)$ is agent $i$'s payment under $M$ when the type profile is $\theta$. One way to
pick $S$ is to discretize the type space and let $S$ be the set of all grid
points. Now what remains to be done is to approximate $C_i(M,\theta)$ using a
linear combination of the variables.

\[C_i(M,\theta)=\frac{\max_{o\in O}\left(\sum_{j\neq i}u_jv_j(\theta_j,o)+a_o\right)
        - \sum_{j\neq i}u_jv_j(\theta_j,o^*)-a_{o^*}}{u_i}\]

Here, $o^*$ is defined as
        \[o^*=\arg\max_{o\in O}\left(\sum_{i=1}^nu_iv_i(\theta_i,o)+a_o\right)\]

        We use the following heuristic to approximate $C_i(M,\theta)$: because the $u_i$ and the $a_o$ are
close to the $u_i^0$ and the $a_o^0$, we will use the $u_i^0$ and the $a_o^0$
to calculate the outcomes mentioned in the above expressions. That is, we
assume that for most type profiles, small perturbation in the parameters will
not change the mechanism outcomes.

\[o^{*0}=\arg\max_{o\in O}\left(\sum_{i=1}^nu_i^0v_i(\theta_i,o)+a_o^0\right)\]
\[o^{0}=\arg\max_{o\in O}\left(\sum_{j\neq i}u_i^0v_i(\theta_i,o)+a_o^0\right)\]

We replace $o^*$ and $o$ using $o^{*0}$ and $o^0$, we have that

\[C_i(M,\theta)\approx\frac{\sum_{j\neq i}u_jv_j(\theta_j,o^0)+a_{o^0}
- \sum_{j\neq i}u_jv_j(\theta_j,o^{*0})-a_{o^{*0}}}{u_i}\]

We use $c_j$ to denote $v_j(\theta_j,o^0)$ and $c_j^*$ to denote
$v_j(\theta_j,o^{*0})$. Both the $c_j$ and the $c_j^*$ are {\em constants}.

\[C_i(M,\theta)\approx\frac{\sum_{j\neq i}c_ju_j+a_{o^0}-\sum_{j\neq i}c_j^*u_j-a_{o^{*0}}}{u_i}\]

We then observe that for any $x$,
$\frac{x}{u_i}=\frac{x}{u_i^0}-\frac{x(u_i-u_i^0)}{u_iu_i^0}$.

We can then rewrite $C_i(M,\theta)$ into:

\[\frac{\sum_{j\neq i}c_ju_j+a_{o^0}-\sum_{j\neq i}c_j^*u_j-a_{o^{*0}}}{u_i^0} -\frac{(\sum_{j\neq i}c_ju_j+a_{o^0}-\sum_{j\neq i}c_j^*u_j-a_{o^{*0}})(u_i-u_i^0)}{u_iu_i^0} \]

The first term is a linear combination of the variables, as $u_i^0$, the $c_j$,
and the $c_j^*$ are all constants.

The second term can be approximated as follows:
\[\frac{(\sum_{j\neq i}c_ju_j^0+a_{o^0}^0-\sum_{j\neq
i}c_j^*u_j^0-a_{o^{*0}}^0)(u_i-u_i^0)}{u_i^0u_i^0} \]

The above is a linear function involving one variable ($u_i$).

Using the above heuristic method, we are able to turn Model 1 into a linear
program involving $n+|O|$ variables. The number of constraints is $n+|O|+|S|$.
Therefore, we can afford reasonable large $n$ and $|O|$, as long as $|S|$ is
not too large.

\section{Zero-Day Exploit Mechanism Design Model}



In this section, we focus on a specific mechanism design setting for the
zero-day exploit model: there are
only two agents: one offender and one defender.

We use $EPO(M)$ to denote the offender's expected payment under mechanism $M$.
We use $EPD(M)$ to denote the defender's expected payment under mechanism $M$.

Let $F$ be the set of all strategy-proof and individually rational mechanisms.

Let $M^*$ be the optimal mechanism that maximizes the expected revenue.
We recall that $ER(M)$ denotes $M$'s expected revenue.

\[M^*=\arg\max_{M\in F}\left(EPO(M)+EPD(M)\right)=\arg\max_{M\in F}ER(M)\]

Let $MO^*$ be the optimal mechanism that maximizes the expected payment
collected from the offender.

\[MO^*=\arg\max_{M\in F}EPO(M)\]

Let $MD^*$ be the optimal mechanism that maximizes the expected payment
collected from the defender.

\[MD^*=\arg\max_{M\in F}EPD(M)\]

Obviously, we have
\[ER(M^*)=EPO(M^*)+EPD(M^*)\le EPO(MO^*)+EPD(MD^*)\]

We introduce the following {\em posted-price} mechanisms.
These mechanisms allow only one agent to make decisions.

\begin{itemize}
    \item Every outcome $o$ is associated with a price $a_o$.
    \item One agent picks the outcome that maximizes her own
        utility.
    \item The other agent makes no decisions and pays $0$.
\end{itemize}

It is without loss of generality to assume that both $MO^*$ and $MD^*$ are
posted price mechanisms. Let $MO^*$ be the mechanism that maximizes the
offender's expected payment. Let $EPO(M,\theta_D)$ be the offender's expected
payment under $M$ when the defender bids $\theta_D$. Because
$EPO(MO^*)=\sum_{\theta_D}P(\theta_D)EPO(MO^*,\theta_D)$, there must exist one
$\theta_D$ so that $EPO(MO^*,\theta_D)\ge EPO(MO^*)$. If we fix the defender's
type to be the said $\theta_D$, then the mechanism faced by the offender is
exactly a posted-price mechanism, and the expected payment of the offender is
at least $EPO(MO^*)$.

For zero-day exploit market, typically the defender has much lower valuation
than the offender. For example, according to~\cite{greenberg2012forbes}, an
exploit that attacks the Chrome browser sells between 80k and 200k for
offensive clients (USD). According to Google's official bug bounty reward
program for the Chrome browser~\cite{guideline2015chrome}, a serious exploit is
priced between 0.5k and 15k. Therefore, $EPD(MD^*)$ is generally much smaller
than $EPO(MO^*)$.

We use $PP(a_0,a_1,\ldots,a_k)$ to denote the posted-price mechanism with the
parameters $a_0$ to $a_k$.
We use $M(u_O,u_D,a_0,a_1,\ldots,a_k)$ to denote the
AMA mechanism with the parameters $u_O$ (for offender), $u_D$ (for defender),
and the $a_t$. If the deciding agent under $PP(a_0,a_1,\ldots,a_k)$ is the
offender, then $PP(a_0,a_1,\ldots,a_k)$ approaches
$M(u_O,1,0,-a_1u_O,\ldots,-a_ku_O)$, when $u_O$ approaches infinity.
If the deciding agent
under $PP(a_0,a_1,\ldots,a_k)$ is the defender, then $PP(a_0,a_1,\ldots,a_k)$
approaches $M(1,u_D,-a_0u_D,\ldots,-a_{k-1}u_D,0)$, when $u_D$ approaches
infinity.
For any posted-price mechanism, there exists an AMA mechanism whose
expected performance is arbitrary close to it. That is, the optimal AMA
mechanism's expected revenue is at least the optimal expected revenue
of posted-price mechanisms.

We solve for the optimal posted-price mechanism for the offender, denoted as
$PP^*$.

\[ER(PP^*)=EPO(M^*)\ge ER(M^*)-EPD(MD^*)\]

Because the optimal AMA mechanism outperforms $PP^*$ (or they have the same
expected revenue), when $EPD(MD^*)$ is small, we have
that the optimal AMA mechanism's expected revenue is close to the
expected revenue of the optimal mechanism $M^*$.


\subsection{Optimal Posted-Price Mechanism}

In this subsection, we discuss how to solve for the optimal posted-price
mechanism.

First of all, we focus on the single-parameter setting~\cite{Guo16:Revenue}.
For presentation purpose, we focus on solving for the optimal posted-price
mechanism where the offender makes decisions.

In a single-parameter setting, an offender's valuation is defined as follows,
assuming her type is $\theta_O$:
\[v(\theta_O,t) = \int_0^t\theta_O c(x)dx\]

$c(x)$ is a fixed function that characterizes the instantaneous valuation of the
exploit by the agent. At time $x$, the instantaneous valuation is $\theta_Oc(x)$.

We use Myerson's standard technique. We assume $\theta_O\in [0,H]$, where $H$
is a fixed upper bound. We assume $\theta_O$'s pdf and cdf are $f$ and $F$,
respectively. We also assume the following expression is monotone
nondecreasing. This is called the {\em monotone hazard rate} condition, which
is satisfied by many common distributions.

\[\phi(\theta_O)=\theta_O-\frac{1-F(\theta_O)}{f(\theta_O)}\]

It turns out that if the above all hold, then the optimal post-price mechanism
simply sells the whole time interval $[0,1]$ as a bundle, with a fixed
take-it-or-leave-it price $p$. If
the agent is willing to afford $p$ to buy the whole interval, then she gets it.
Otherwise, she gets nothing and pays nothing. The optimal mechanism $PP^*=PP(0,\infty,\infty,\ldots,\infty,p)$.

When $k$ is small, we have another algorithm for solving for the optimal
posted-price mechanism, and under this algorithm, we can drop the
single-parameter assumption.

Let $PP(a_0,a_1,\ldots,a_k)$ be the optimal posted-price mechanism. It is
without loss of generality to assume that for any $i<j$, if both $a_i$ and
$a_j$ are finite, then $a_i<a_j$. Otherwise, the outcome $i$ is never chosen,
and we can set $a_i$ to be infinite. It is also without loss of generality to
assume that for any $a_i$, either it is infinite (meaning that this outcome is
not allowed), or it must satisfy the following condition:

\[ \exists \theta_O, v(i,\theta_O)-a_i=\max_{j<i}v(j,\theta_O)-a_j\]

Here, $v(t,\theta_O)$ is the offender's valuation for outcome $t$ when her type
is $\theta_O$. The above condition basically says that there exists a type for
the offender, if we increase $a_i$ just a bit, then it would force the offender
to choose an earlier outcome than $i$. If the condition is not true, then we
can safely increase $a_j$. By doing so, we can charge more for those types that
choose $j$. We can also charge more if under some types, the offender chooses a
later outcome, which also means that more payment will be collected.

Based on the above condition, if we know the $a_j$ for $j<i$ and $\theta_O$,
then we can calculate $a_i$. We already know that $a_0=0$. To
calculate $a_1$, we can go over all $\theta_O$, possibly by discretizing the offender's
type space. Then, to calculate $a_2$, we can go over all $\theta_O$ again. We
do this for every $i$. Let $N$ be the number of types in $\Theta_O$ after the
discretization. The total number of iterations is then $N^k$.

\section{Evaluation}

As mentioned earlier, according to~\cite{greenberg2012forbes}, an exploit that
attacks the Chrome browser sells for at most 200k for offensive clients
(USD). According to Google's official bug bounty reward program for the Chrome
browser~\cite{guideline2015chrome}, a serious exploit is priced for at most 15k.

We start with the following setting, which is based on the
numbers above. There are two agents.
The offender's valuation function is \[v(\theta_O,t) =
\int_0^t\theta_O(1-x)dx\] $\theta_O$ is drawn uniformly at random from $U(0,
400)$. That is, the offender's valuation for the whole time
interval $[0,1]$ is at most $200$.

The defender's valuation function is
\[v(\theta_D,t) = \int_t^1\theta_Dxdx\]
$\theta_D$ is drawn uniformly at random from $U(0, 15)$. That is, the
defender's valuation for the whole time
interval $[0,1]$ is at most $15$.

The above valuation functions satisfy all the conditions needed for the
single-parameter model. So $MO^*$ simply sells the whole interval to the
offender for a fixed price $p_O$, and $MD^*$ simply sells the whole interval to
the defender for a fixed price $p_D$.

\[ p_O = \arg\max_{p\le 200}pP(v(\theta_O,1)\ge p) = \arg\max_{p\le 200}p\frac{200-p}{200} = 100 \]
\[ EPO(MO^*) = 50\]
\[ p_D = \arg\max_{p\le 15}pP(v(\theta_D,1)\ge p) = \arg\max_{p\le 15}p\frac{15-p}{15} = 7.5\]
\[ EPD(MD^*) = 3.75\]

Therefore, $ER(M^*)$ is at most $53.75$. We pick $k=10$ and $\epsilon=0.01$.
We use the VCG mechanism as the initial solution. The VCG mechanism's expected
revenue is $6.9$, which is very far away from the upper bound. Our technique
starts from the VCG mechanism, and at the end produces a mechanism whose
expected revenue equals $50.6$, which is very close to the upper bound $53.75$.
That is, in this case, the optimal AMA mechanism's expected revenue is close to the optimal
mechanism's expected revenue.

As demonstrated in our analysis, we have the above phenomenon if the defender's
valuation is insignificant compared to the valuation of the offender. We then
investigate an example where the defender's valuation is much higher. We
change it so that the defender's type is drawn from $U(0,150)$ instead of
$U(0,15)$. Now the upper bound for $ER(M^*)$ is $87.5$. Our technique produces
a mechanism whose expected revenue equals $57.9$. This time, the achieved value
is not close to the upper bound.

\section{Conclusion} Optimizing within the affine maximizer auctions (AMA) is
an effective approach for revenue maximizing mechanism design. We proposed a
linear programming based heuristic for optimizing within the AMA family. We
applied our technique to revenue maximizing mechanism design for zero-day
exploit markets. We showed that due to the nature of the zero-day exploit
markets, with one offender and one defender, our technique generally produces a
near optimal mechanism.

\bibliographystyle{splncs03}
\bibliography{zeroday2}
\end{document}